\documentclass[floatfix,preprintnumbers,superscriptaddress,showpacs,
               showkeys,prb,preprint]{revtex4}
\usepackage[version=3]{mhchem} 
\usepackage{graphicx}

\begin{document}
\bibliographystyle{apsrev}
\preprint{2010/TiO2Si}

\title{Characterization of silicon thin overlayers on \\
rutile \ce{TiO2} (110)-(1x1)}

\author{J. Abad}
\email{jabad@um.es}

\affiliation{
Departamento de Fisica, Centro de Investigacion en Optica y Nanofisica-CIOyN
(Campus Espinardo), Universidad de Murcia, E-30100 Murcia, SPAIN
}

\author{C. Gonzalez}
\author{P.L. de Andres}
\author{E. Roman}
\affiliation{
Instituto de Ciencia de Materiales de Madrid (CSIC)
E-28049 Cantoblanco, Madrid, SPAIN
}

\date{\today}

\begin{abstract}
Silicon thin films for coverages ($\theta$) between 0.3 and 3 monolayers have 
been grown on rutile \ce{TiO2}(110)-(1x1) at room temperature and studied by 
x-ray and ultra-violet photoelectron spectroscopies, Auger electron spectroscopy, 
and low energy electron diffraction (LEED).  
A clear evidence of a strong \ce{Si}/\ce{TiO2} interaction consistent with the high
affinity of O for Si has been found. The Ti cations on the substrate are reduced, 
while the Si film is oxidized, yielding \ce{SiO2} and a mixture of silicon suboxides.  
Neutral Si atoms are observed at a coverage of 3 monolayers.  
At the interface region we observe the formation of cross-linking Ti-O-Si bonds. 
The thin Si overlayer strongly attenuates the $(1 \times 1)$ LEED pattern from 
the substrate. Finally, thermal annealing results in the improvement of the \ce{SiO2} 
stoichiometry, but the surface order is not recovered. 
Using ab-initio density functional theory we have obtained optimum geometrical
configurations and corresponding density of states for $\frac{1}{3} \le \theta \le 1$ 
monolayers of Si adsorbed on the $1 \times 1$ two-dimensional unit cell.
\end{abstract}






\maketitle

\section{Introduction}

SiO$_{x}$-TiO$_{2}$ ($0 \le x \le 2$) mixed materials are interesting for  many
technological areas ranging from heterogeneous catalysis to semiconductor industry.
Materials based on mixtures of titanium and silicon have been extensively studied 
due to their catalytic properties.\cite{gao99,1,2} 
These oxides are good model systems in catalysis providing also a convenient
support for metal catalyst preparation and  to study the strong metal-support
interaction (SMSI).\cite{3} 
Furthermore, the formation of Ti-O-Si cross-linking bonds strongly modifies 
the electronic structure of the interface, yielding unique catalytic properties.\cite{4}
Titanium-silica films have also received much attention in optical applications due 
to their high thermal and chemical stability and controlled refractive index.\cite{5} 
In the field of the renewable energy generation \ce{SiO2}-\ce{TiO2} thin films are 
used as working electrodes for dye-sensitized solar cells.\cite{6} 
Furthermore, in micro-electronics new generations of metal-oxide-semiconductor
(MOS) devices demand a reduction of the thickness of the insulating layer. 
Among the various possible insulators, TiO$_{2}$ has received an 
increasing amount of attention in recent years.\cite{7,8} 

\begin{figure}
\begin{tabular}{cc}
\includegraphics[width=0.5\columnwidth]{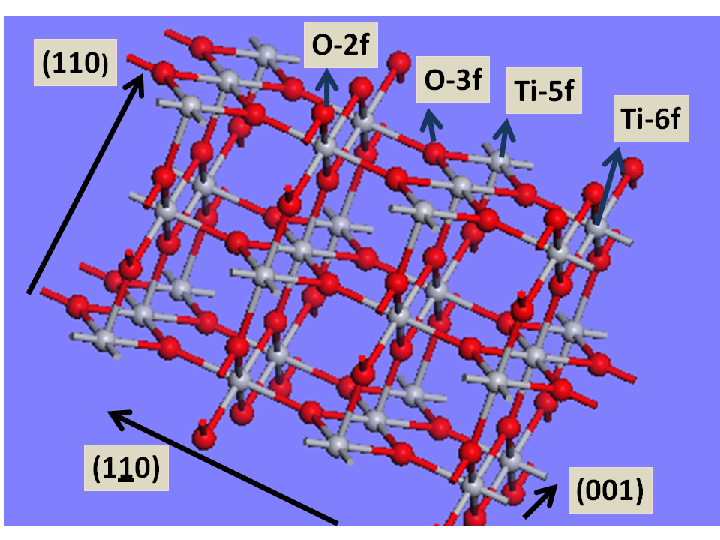}  &
\includegraphics[width=0.5\columnwidth]{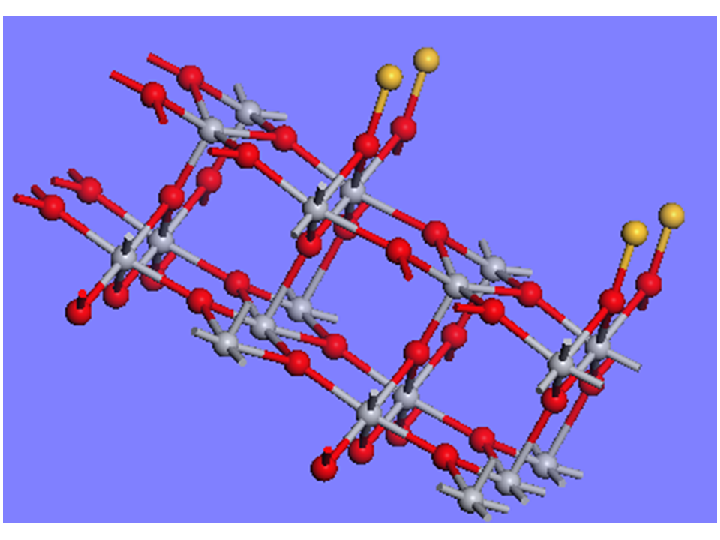}  \\
\end{tabular} 
\caption{(Color online) 
(a) Left panel: Model for clean \ce{TiO2} -1x1 (110).  The actual slab is made of five layers (only 
the two topmost layers 
are shown for the sake of clarity). 
(b) Right panel: Low coverage optimum DFT model for
\ce{Si}/\ce{TiO2} (110)-(1x1). The preferred adsorption site for Si is located over the two-fold
coordinated bridging oxygens.
}
\label{fgr:1}
\end{figure}

An important issue in the semiconductor or oxide-oxide systems is the sharpness 
of the interface and the presence of suboxides of well-defined stoichiometry.  
The growth of SiO$_{2}$ films on TiO$_{2}$ has been studied by 
Barranco et al.\cite{9}, concluding that for low coverages ($\theta$) \ce{SiO2} grows 
in a layer by layer mode. 
Finally, the inverse system, i.e. TiO$_{2}$ on SiO$_{2}$ substrates\cite{10,11,12} ,
as well as the oxidation of Ti films deposited on mono and multilayer 
\ce{SiO2}/Mo(112) surfaces, are interesting in their own and have been reported 
in the literature too.\cite{13} 
The \ce{TiO2}(110)-(1x1) is the most studied rutile surface owing to its stability, 
a well established experimental procedure to obtain it, and the accurate 
determination of its composition and geometry.\cite{dieboldSSR48,cabailh07}
It represents  the bulk terminated stoichiometric surface, where two kind of Ti atoms
can be found; along the [001] direction, rows of sixfold coordinated Ti atoms (Ti-6f)
alternate with fivefold coordinated Ti atoms (Ti-5f). Similarly, two different classes of 
O atoms are created: in the main surface plane there are threefold coordinated 
in-plane O (O-3f), and twofold coordinated (O-2f) protruding bridging O atoms 
(see Fig.~\ref{fgr:1}(a), left panel). 
A small density of disordered point defects characterizes this structure. 
These have been attributed to vacancies in the bridging O rows produced during 
the sample preparation, sputtering and annealing in ultra high vacuum 
(UHV).\cite{dieboldSSR48,18}  

In this paper, we present experimental and theoretical work for Si thin films at low
(0.3 ML), and high coverage (3 ML), on the \ce{TiO2}(110)-(1x1) stoichiometric
surface. Although experiments show some disorder in the interface, our
calculations have been performed for a $1 \times1$ 2D unit cell, and validated
by comparing with similar results for a sparse $2 \times 2$, more appropriated 
to describe disorder. 
As long as the Si coverage remains low (early stages of growth, one Si/unit cell) 
our results indicate that interaction of Si with O dominates, making the initial energy
landscape rather insensitive to the use of a $1 \times1$ or a $2 \times 2$ unit cell.
For the initial stages of growth our theoretical calculations predict several adsorption
sites within a narrow energy window, explaining the existence of disorder at room
temperature by random population of these. Once the available bonds to O have 
been saturated (two Si/unit cell), it is most logical for a feasible theoretical model to
follow the $1 \times 1$ periodicity imposed by the substrate.
It is only when the film grows thicker (more than three Si/unit cell) that diffusion of
interstitials and epitaxial stresses become important, and a more complex model
beyond our computational means should be consider to understand the amount of
disorder at the interface with a thick SiOx film. This is beyond the scope of our
theoretical approach. However, based in the available experimental and theoretical
work for the regime of interest here, we can conclude that at the early stages of
growth the Si film is oxidized, whereas the Ti cations are reduced, and Ti-O-Si 
cross-links bonds are formed at the interface. For coverages larger than 3 ML 
neutral Si atoms are experimentally observed. We have also found that the 
\ce{SiO2} stoichiometry is improved in the 3 ML thick film upon thermal annealing,
although the surface order as monitored by the low-energy electron diffraction
(LEED) pattern is not improved by this heating treatment.
According to our density functional theory (DFT) calculations we propose models
containing from one (low coverage) to three (high coverage) Si atoms per 
TiO$_{2}$ (110)-(1x1) two-dimensional unit cell (adsorption of three Si atoms
completes 1 ML, since a fourth Si atom cannot be directly bonded to the substrate).
The theoretical results will be compared with the experimental data obtained for
coverages under and above the monolayer. 

\section{Experimental}

Experiments were carried out in an UHV chamber, equipped with Auger Electron
Spectroscopy (AES), X-Ray Photoelectron Spectroscopy (XPS), Ultraviolet
photoelectron Spectroscopy (UPS), LEED, 
and argon ion sputtering; details of the system have been published  
elsewhere.\cite{abad04} 
The ultimate pressure was $2 \times 10^{-10}$ mbar. The \ce{TiO2} (110) rutile 
single crystals ($10 \times 10 \times 1$ mm$^{3}$ with both sides polished) used 
in this study were provided by PI-KEM Ltd. UK. 
The \ce{TiO2} single crystal was bulk reduced and surface cleaned by repeated
cycles of Ar$^{+}$ bombarding (typically $0.5-3.0$ keV), followed by annealing 
($600^{\circ}$C, 60 min) until no impurities were detected by AES and XPS. 
After several of these cycles the colorless intrinsic \ce{TiO2} turned to blue. 
As it has been documented on the literature a crystal of light blue color has enough bulk defects to avoid charging effects
during the photoemission experiments and a deep blue color is related 
to $\approx 10^{19}$ oxygen vacancies per cm$^{3}$.\cite{cronomeyer52}
The \ce{TiO2} (110)-(1x1) surface was prepared by annealing up to 
$450^{\circ}$C in \ce{O2} atmosphere ($6.5 \times 10^{-7}$ mbar)
during 90 minutes,
and by subsequent cooling of the sample in \ce{O2} atmosphere during 30 minutes.\cite{20,21} 
It has been reported in the literature how annealing at lower temperatures
($250^{\circ}$- $390^{\circ}$ C) could lead to rosette 
structures.\cite{dieboldSSR48}
With the techniques used in this manuscript we cannot fully 
ascertain the absence
of these structures even if after our annealing schedule we have observed
a sharp $(1 \times 1)$ LEED pattern, corresponding to a 
well ordered surface.
However, we notice that our annealing procedure goes to higher temperatures
than the ones reported for the formation of rosettes, 
making more likely that the current surface would be free of these. 
We remark that techniques used in this work do not provide 
direct access to detailed information about the number or
distribution of defects on the surface, as for example the
presence of oxygen adatoms, 
but the annealing to higher temperatures usually work to
improve matters, in particular it has been reported that
annealing to higher temperatures than 330$^{\circ}$ C helps to get rid
of these defects.\cite{dieboldSSR48}
Pure Si was deposited on this \ce{TiO2} (110)-(1x1) surface at 
room temperature (RT) using a home-made e-beam evaporator with a 
high purity Si rod as evaporation source. 
The rate of deposition was determined by evaporating Si 
on a polycrystalline Au substrate and determining the coverage
by AES.\cite{22} 
In this study only two cases have been considered: 0.3 ML and 3 ML.
Base pressure during the ulterior annealing of the Si 3 ML thick film has been monitored to be always
below $10^{-9}$ mbar to avoid contamination by traces of \ce{H2O} or other residual gases
(the sample holder was thoroughly degassed by going to higher temperatures than the ones
used for Si deposition).
The chamber is equipped with two non-monochromatic X-Ray sources, Al K$_{\alpha}$ (1486.6 eV) and Mg K$_{\alpha}$ (1253.6 eV), a He-discharge lamp (21.2 and 40.8 eV photon energy), and an electron gun of 3 keV energy. UPS, AES and XPS spectra were recorded using a hemispherical analyzer. The overall resolution (source and analyzer) is 1.0 eV for AES, 0.8 eV for XPS and 0.3 eV for UPS measurements. The contribution of the Mg K$_{\alpha}$ satellite lines and a Shirley type background were subtracted from the experimental spectra before the XPS data were analyzed.\cite{23,24} The Ti 2p, Si 2p and O 1s peaks were fitted using Voigt curves.  AES data were acquired in the integrate mode N(E). 

\section{Theoretical}

Ab-initio density functional theory calculations have been performed 
using soft pseudopotentials\cite{27} and a plane-waves basis with a cutoff energy between 350 and 375 eV.
The exchange and correlation functional has been computed in a generalized gradients approximation suggested by Perdew, Burke and Ernzerhof,\cite{26} and the Brillouin Zone  has been sampled on a $8 \times 4 \times 1$ Monkhorst and Pack mesh.\cite{52}  
Actual calculations have been performed using the CASTEP program.\cite{25}
To characterize the electronic distribution in the near-surface region as accurately as possible
we use a five layers slab to model the \ce{TiO2} (110)-(1x1)  surface, with
one or two layers at the bottom of the slab kept frozen in bulk positions. 
All the calculations have been done at  $T=0$ K,  relaxing the system until the total energy changed 
less than $10^{-4}$ eV, and the maximum force on any ion was less than 0.01 eV/{\AA}.
The relaxed geometry has been passed to a different DFT program working with a localized basis set
(FIREBALL\cite{28}).
Wave-functions are now expanded into a basis of numerical atomic orbitals with a finite spatial range 
(spatial cutoffs, see Table~\ref{tbl:1}), 
yielding  immediate access to the density of states (DOS) 
projected on different atoms. 
This reduced basis represents a simpler approach with respect to the use of plane-waves, 
but it is known to produce good results at a moderated computing cost.\cite{29}

\section{Results and discussion}

\subsection{Si adsorption site}

\begin{figure}
\begin{tabular}{cc}
\includegraphics[width=0.60\columnwidth]{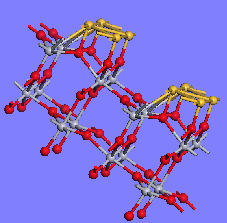}  &
\includegraphics[width=0.36\columnwidth]{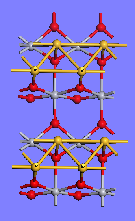}  \\
\end{tabular}
\caption{(Color online)
Side and front views for two Si adatoms
on  \ce{Si}/\ce{TiO2} (110)-(1x1). 
Si shows a tendency to oxidize making zigzag chains running
along the (001) direction.
}
\label{fgr:2}
\end{figure}

\begin{figure}
\begin{tabular}{cc}
\includegraphics[width=0.60\columnwidth]{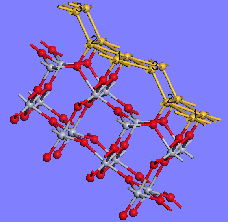}  &
\includegraphics[width=0.36\columnwidth]{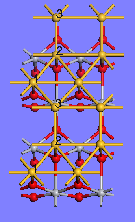}  \\
\end{tabular}
\caption{(Color online)
Side and front views for the complete first Si monolayer
(three Si adatoms adsorbed on \ce{TiO2}(110)-(1x1)).
Every Si is located near atop positions with respect to
oxygens in 1x1 unit cell, at distances $d \approx 1.8$ {\AA},
except the last one that goes up to a Si-O distance of
$2.6$ {\AA} to release the stress of Si(2)-Si(3) bond
( $d \approx 2.7$ {\AA}).
}
\label{fgr:3Si}
\end{figure}

So far, there is no direct information about the exact geometry of Si
on the \ce{TiO2}(110)-(1x1) surface. 
Since LEED has not been helpful trying to solve this question, we
have used ab-initio density functional theory to 
optimize the adsorption site of a single Si adatom (Si-1) on the 
unit cell:
seven different initial positions taking into account all the symmetry sites 
and spanning uniformly the 2D 1x1 unit cell have been
investigated.
The absolute minimum, corresponding to the lowest total energy, 
has been found for Si interacting
directly with a twofold coordinated bridging oxygen (O-2f) 
at a bonding distance of $d=1.65$ {\AA} (Fig.~\ref{fgr:1}(b), right panel). 
The closest titanium (Ti-6f) is located at $3.44$ {\AA}, far enough to preclude an important interaction, 
and the nearest next oxygen (O-3f) is at $3.66$ {\AA}.
This adsorption site is quasi-degenerated in energy ($0.06$ eV) with the second-best stable position where
Si sits nearly equidistant between one O-2f and one O-3f oxygen atoms 
($d=1.8$ {\AA}) and is located at $2.9$ {\AA} from the nearest Ti-6f. 
The existence of different adsorption sites inside a narrow
window of energies explains the tendency of these layers to
show disorder when grown at room temperature.
Similar calculations performed on a $2 \times 2$ unit cell 
(this slab has been made with three tri-layers) support the same picture
with minor changes: for the optimum adsorption position the Si-O distance
decreases to $1.56$ {\AA}, and the energy difference with the second
best near equivalent position is increased to $0.11$ eV. 
Finally, the most disadvantageous 
{\it meta-stable} site 
we have found ($+0.22$ eV) locates Si as the vertex of a pyramid  
remaining at $d \approx 1.9$ {\AA} from two O-2f and one O-3f, and  $2.7$ {\AA} from the Ti-6f located directly below.
These configurations model the low-coverage scenario ($\theta=0.3$ ML).
To investigate the high coverage regime we add an extra Si-atom to our previous optimum geometry (Si-2).
Two different stable configurations have been found with an energy difference of $0.25$ eV. 
The most stable one (Fig.~\ref{fgr:2}) locates the second Si-atom directly above the in-plane three-fold coordinated O atom, 
at $2.67$ {\AA} from the first Si atom; this is larger than the typical Si-Si distance in its crystalline form ($2.35$ {\AA}), 
but quite comparable to the Si-Si distance in the presence of oxygen, e.g. $2.665$ {\AA} in stishovite,
a dense tetragonal form of \ce{SiO2}  with the rutile structure.
Therefore, the role of oxygen in the Si-Si interaction seems to be clearly established. 
We notice that in this configuration the second Si atom is not accommodated in the second best position for the 
adsorption of a single Si (between two oxygens). 
The optimum structure forms a Si zigzag pattern running along the (001) direction over the bridging O-2f (Fig.~\ref{fgr:2}), 
while the second best candidate locates the adatoms in a linear disposition along the $(1\overline{1}0)$ direction. 
Finally, we saturate the Si layer by locating a third Si atom 
(Si-3, $\theta=1$) in a symmetrical position between two zig-zag lines.
Relevant distances for the optimum configuration are given in Table~\ref{tbl:2}, 
where  we conclude that Si-1 and Si-2 remain bound to the surface through 
oxygen atoms while
Si-3 forms a bond to Si-2 ($d=2.45$ {\AA}, comparable to the Si-Si distance in crystalline silicon).
At the same time, Si-1 to Si-2 and Si-1 to Si-3 distances
remain large, $\approx 2.6-2.7$ {\AA} (Fig.~\ref{fgr:3Si}),
making difficult for the last silicon arrived at the surface (Si-3) 
to interact with the oxygen directly below ($d \approx 3.7$ {\AA}).
We find a metastable configuration ($+0.5$ eV) with a Si-O distance of 
$\approx 1.8$ {\AA} and a 
Si(3)-Si(2) bond length of $\approx 2.3$ {\AA}. 
The difference in the total energy is important enough to
make us to conclude that the short Si-Si distance in the presence of oxygen is unfavorable and cannot be compensated by the increased Si-O interaction.
These results are summarized in Table~\ref{tbl:2}.


\subsection{XPS}
\begin{figure}
\includegraphics[width=0.95\columnwidth]{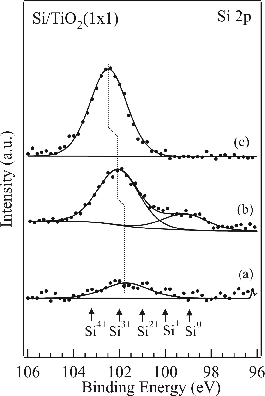}  
\caption{XPS Si 2p spectra for: (a)  0.3 ML covered surface,  (b) 3 ML, 
and (c) the 3 ML Si covered surface after 30 minutes annealing at $460^{\circ}$ C. 
The dotted line shows the shift to higher binding energies.}
  \label{fgr:4}
\end{figure}

\begin{figure}
\includegraphics[width=0.95\columnwidth]{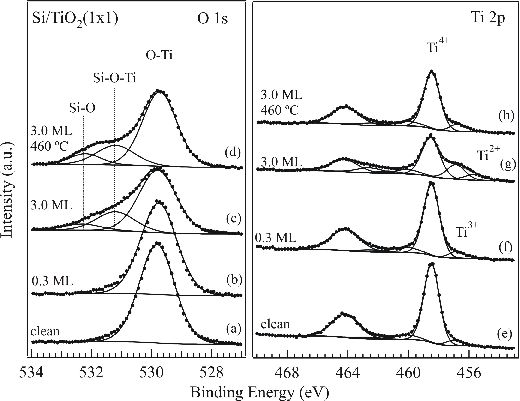}  
\caption{XPS 
O1s (a) --left panel-- and 
Ti 2p (b) --right panel-- spectra for:
(1x1) clean surface
and Si covered surfaces 
as a function of Si dose and after annealing during 30 minutes at 
$460^{\circ}$ C.
}
  \label{fgr:5}
\end{figure}

To characterize the nature of the \ce{Si}-\ce{TiO2} 
interface, we have studied the XPS of Si 2p, Ti 2p and O 1s core 
level spectra. 
Fig.~\ref{fgr:4} compares the XPS Si 2p spectra recorded 
at 0.3 ML and 3 ML of Si coverage, and the 3 ML after annealing. 
The sub-monolayer deposits are characterized by a single Si 2p line
centred at 101.8 eV, while two peaks at 102 and 99.2 eV characterize 
the experimental spectra of the multilayer film. 
After annealing of the multilayer film, one peak centred at 102.5 eV 
has been again produced. 
The values of the Si oxidised states reported in the literature are 
indicated with arrows on the figure 
(\ce{Si}$^{4+}$ = 103.2 eV, 
\ce{Si}$^{3+}$= 102.0 eV, 
\ce{Si}$^{2+}$= 101.0 eV, 
\ce{Si}$^{1+}$=100.0 eV,
and Si$^{0}$ = 98.9 eV).\cite{37} 
In general, it is assumed that the presence of 
Si$^{n+}$ states (n = 1, 2, 3 and 4) is due to silicon atoms bonded 
to nearest neighbour O atoms. 
In the case of the Si-O bond a shift of the Si 2p level of 1.05 eV 
is seen in agreement with previous reported results.\cite{38} 
Therefore, at a coverage of 0.3 ML silicon is oxidized in an 
average oxidation state of Si$^{3+}$. 
On the other hand, at a coverage of 3 ML another state corresponding 
to Si$^{0}$ at the binding energy 99.2 eV appears, 
in agreement with our own AES spectra (not shown), 
and the theoretical results discussed below. 
The analysis of the Si 2p core level peaks show around a 20\% 
of Si-Si bonds and 80\% 
of Si-O bonds on the surface. 
After annealing the 3 ML Si-dosed surface to 
$460^{\circ}$ C for 30 min, the low binding energy components 
disappear completely and only one peak located at 102.5 eV is observed, 
revealing that annealing increases the oxidation state of Si. 
The dotted line in Fig.~\ref{fgr:4} shows the shift to higher 
binding energies indicating an increase in the oxide stoichiometry, 
always below the value assigned for bulk \ce{SiO2}, 103.2 eV. 
This result indicates a great concentration of Si$^{3+}$ oxide, 
but we cannot rule out the presence of sub-oxides SiO$_{x}$ species 
with $1 \le x \le 2$,
and \ce{SiO2}.\cite{37}
We have used AES to gather further information on the interface,
a technique that has been frequently used in the past to study the Si-O 
bonding at different Si interfaces. We have studied the Si LVV AES 
transition to measure data for three different cases: (a) the clean 
(1x1) surface covered with 0.3 ML of Si, (b) the same surface with 3 
ML of Si, and (c) the 3 ML Si covered surface after 30 minutes annealing 
at $460 ^{\circ}$C. The Si LVV shows chemical shifts that may be used to 
distinguish silicon oxides and silicon, but for low Si coverage the AES 
intensities are too similar to the clean surface to allow a reliable analysis
of these spectra. When 3 ML of Si are deposited, the Si LVV transition 
exhibits a peak near 64 eV, and a broad feature between 77 eV and 90 eV. 
Therefore, even if a quantitative analysis is not possible, AES qualitatively
agrees with XPS for Si coverages between 0.3 and 3 ML. After annealing the 
3 ML film at $460 ^{\circ}$C during 30 min. a broad peak at the \ce{SiO2} 
kinetic energy is found, which we take as an indication of the progressive
build-up of stoichiometric \ce{SiO2}. The Si oxidation might be produced by 
either diffusion of O from \ce{TiO2} bulk to the surface, or by diffusion 
of surface Ti into the bulk during annealing. 
Fig.~\ref{fgr:5} shows the XPS spectra of the O 1s (left part) 
and Ti 2p (right part) core levels. 
The spectra have been fitted using Voigt functions. 
The O 1s core level spectrum for the clean (1x1) surface can be 
fitted to a single peak at 529.8 eV 
(Fig.~\ref{fgr:5}(a)), corresponding to O-Ti bonds.\cite{39} 
We notice that for 0.3 ML there is a small increase 
in the high binding energy region at 531.2 eV (Fig.~\ref{fgr:5}(c)).

It is questionable to assign this feature to Si-O-Ti 
cross-linking bonds between the SiO$_{x}$ overlayer and the 
\ce{TiO2} substrate, since  fitting  this feature we obtain a value of 
about 3\% of the total O 1s area, which is within the experimental error bars in the fitting procedure. 
However, at a Si coverage of 3 ML (Fig.~\ref{fgr:5}(d)) a tail in the high binding energy is better defined and a best fit for the O 1s spectrum is obtained  with three components at binding energies of 529.9 eV (O-Ti), 531.3 eV  (Si-O-Ti), and  532.2 eV (O-Si). These new components indicate the coexistence of different O chemical bonds, because Si and Ti cations compete strongly for lattice O ions, resulting in the formation of mixed-oxide states at the interface. 
The (Si-O-Ti) component is assigned to Si-O-Ti cross-link bonds between the SiO$_{x}$ overlayer and the 
\ce{TiO2} substrate, as occurs in the case of Si on the (1x2) reconstructed surface.\cite{abad06} 
The growth of the tail towards a higher binding energy can be explained
if two Ti-O-Ti bonds are replaced by a single Ti-O-Si bond, namely
because the number of Ti-O bonds decrease due to the formation of Si-O
bonds. As Si is more electronegative than Ti, this difference should
be responsible for the binding energy shift observed in Fig.~\ref{fgr:5}
(left panel). 
This Ti-O-Si bond is supported by results already reported  for the 
\ce{TiO2}-\ce{SiO2} oxide systems, \cite{45,46} 
as well as for \ce{TiO2} films deposited on  \ce{SiO2}.\cite{10,11}
The Ti 2p core level spectrum of the clean (1x1) surface (Fig.~\ref{fgr:5} (e)) is characterized by the splitting of 
2p$_{3/2}$ and 2p$_{5/2}$. The 2p$_{3/2}$  is fitted by two components:
The most intense one is located at 458.5 eV binding energy corresponding to Ti$^{4+}$ cations.\cite{abad04,39,40} 
The second one is centred at a binding energy of 456.8 eV 
($\Delta E = 1.7$ eV), 
which reveals the presence of Ti$^{3+}$ cations,\cite{abad04,21,41,42} 
and is related with O vacancies present on the (1x1) stoichiometric surface.\cite{18} 
In accordance with earlier work,\cite{oku99,abad04}
we have introduced a satellite component (BE 460 eV)
to get a good fit of the Ti 2p peak.
From the area of the Ti-2p$_{3/2}$ components 
(given in percent)
the reduction of the surface, 
$\delta$, 
is determined to be $0.01 \times [Ti^{3+}] + 0.02 \times [Ti^{2+}]$. 
This parameter characterizes the average oxidation or reduction of the 
surface substrate.\cite{idris94,barteau96,mostefa99,abad04} 

For the stoichiometric $1 \times 1$ surface,
the values of the  surface reduction, $\delta$, 
and the O/Ti ratio obtained from 
$\frac{\ce{O}}{\ce{Ti}} = 2-\frac{\delta}{2}$,  
are found to be $0.03$ and $1.99$, respectively. 
At a Si coverage of 0.3 ML, (Fig.~\ref{fgr:5} (f)) 
these parameters become $0.04$ and $1.96$, 
which are not significantly different from those of the stoichiometric 
$ 1 \times 1$ surface. 
At a coverage of 3 ML the results of the Ti-2p$_{3/2}$ core level 
(Fig.~\ref{fgr:5}(g)) 
are fitted by three components at binding energies of  
458.5 eV (Ti$^{4+}$),
456.8 eV (Ti$^{3+}$), and 
455.4 eV (Ti$^{2+}$).\cite{abad04} 
The surface reduction, 
$\delta$, and the O/Ti ratio are 0.40 and 1.80, respectively. 
These values agree well with related results reported earlier for the 
$1 \times 2$ reconstructed surface of 
$0.37$ and  $1.82$ at 2 ML and  $0.52$ and $1.74$ at 4 ML.\cite{abad06} 
Notice that these parameters show a clear dependency with the Si coverage,
regardless of the \ce{TiO2} reconstructed surface. 

The presence of Ti$^{3+}$ and Ti$^{2+}$ states indicates that Si induces a strong reduction of the substrate atoms. 
There is also a strong increase of the $\delta$ parameter with the Si coverage, 
about  one order of magnitude larger at 3 ML than the values corresponding to the clean surface. 
The reaction taking place at RT on the interface is given by:
$$
\ce{TiO2} + \ce{Si} \rightarrow \ce{TiO}_{x} + \ce{SiO}_{y}
$$
with $x  \le  2$ and $0 \le y \le 2$.  

After annealing at $460^{\circ}$ C during 30 minutes the O 1s spectrum (Fig.~\ref{fgr:5} (d)) 
shows a slight intensity decrease in the Ti-O-Si component located at 531.2 eV. 
This implies that annealing breaks Ti-O-Si bonds and  O forms new bonds to Ti or Si atoms. 
Ingo et al.\cite{46} have reported a decrease of this component after heating a mixture of \ce{TiO2}-\ce{SiO2} 
oxides at $600^{\circ}$ C in air, proving the creation of new Ti-O and Si-O bonds. 
At the same time, the analysis of Ti 2p core level (Fig.~\ref{fgr:5} (h)) shows that after annealing the component 
associated with the Ti$^{2+}$ state disappears completely. 
The surface reduction $\delta$ and the O/Ti ratio 
change from $0.40$, and $1.80$, to $0.09$, and $1.96$
before and after the annealing respectively. 
This is a clear indication that Ti cations are oxidized by the annealing.
It is noteworthy to notice that the density of defects present in 
the annealed surface is similar to the one found on the surface 
reconstruction (1x2). However, as discussed below, after annealing 
the surface does not show a clear LEED pattern.


\subsection{UPS and DOS}

\begin{figure}
\includegraphics[width=0.95\columnwidth]{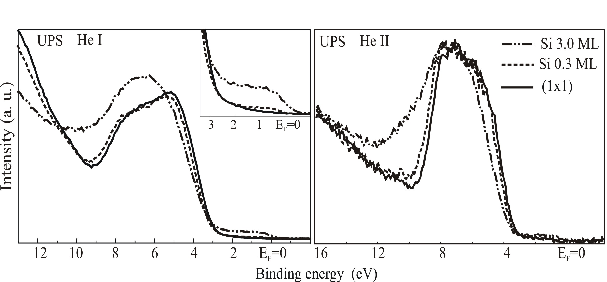}  
\caption{
(a) Left panel: UPS He I spectra (inset shows the band gap region). 
(b) Right panel: UPS He II spectra for the clean (1x1) surface and Si covered surfaces
}
  \label{fgr:6}
\end{figure}

\begin{figure}
\includegraphics[width=0.95\columnwidth]{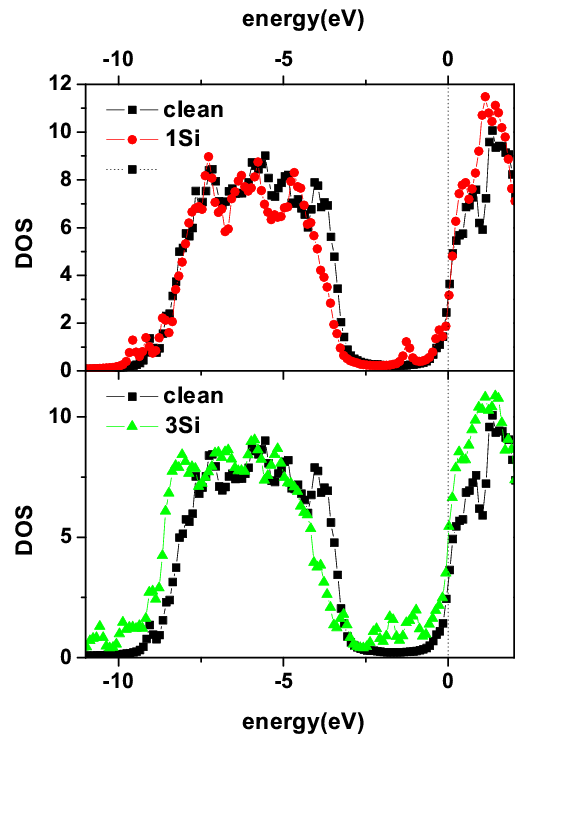}  
\caption{
(Color online) DOS of the most stable structures with one (upper panel) and three (lower panel) Si atoms, 
compared with the clean \ce{TiO2}-surface. The main effect of Si on the DOS is to inject new states in
the lower part of the spectra ($\approx -8.5$ eV) and in the gap. 
}
  \label{fgr:7}
\end{figure}

\begin{figure}
\includegraphics[width=0.95\columnwidth]{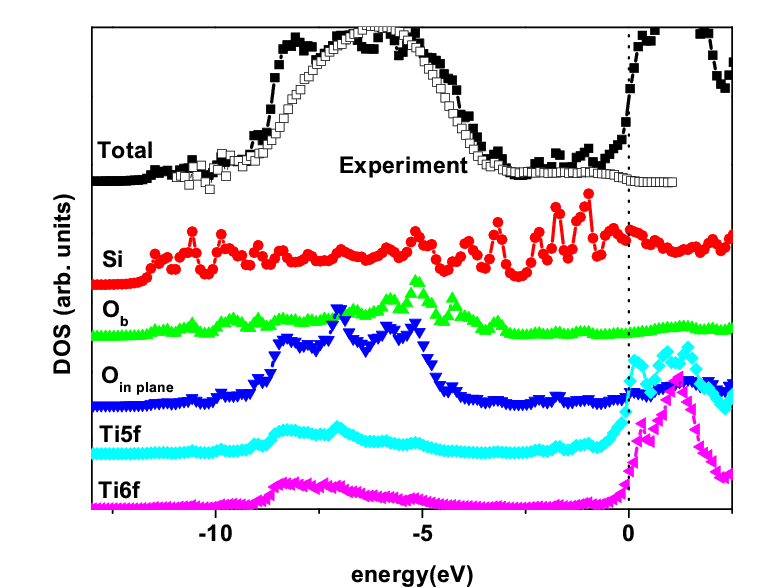}  
\caption{(color online): Black squares compare the total DOS of the three Si-atoms and the experimental 
UPS measurements for the 3ML case. Red circles show the Si-atoms contribution to the DOS. 
The other lines are the same for the bridging O (O-2f), in plane O (O-3f), and the 5-fold and 6-fold coordinated 
Ti atoms (mainly contributing to the conduction band, not measured in the experiment). 
}
  \label{fgr:8}
\end{figure}

In this section, we shall analyze the experimental UPS spectra by comparing
with the DOS calculated theoretically for 
the structures described above 
(Figs.~\ref{fgr:1}b, ~\ref{fgr:2}, and ~\ref{fgr:3Si}).
Fig.~\ref{fgr:6} shows the valence-band spectra obtained by 
UPS (He I --left panel--,  He II --right panel--) 
for the (1x1) clean surface, and for surface coverages of 0.3 ML and 
3.0 ML, respectively. 
Our theoretical calculations show that the \ce{TiO2}(110)-(1x1) 
valence-band structure is mainly due to O 2p orbitals, with a slight
hybridization between the O 2p and Ti 3d orbitals.
At Si 0.3 ML coverage the line shape of the UPS spectrum does not change 
with respect to the pristine surface, 
although a valence band shift of $0.2$ eV towards higher binding energies 
is observed. 
This displacement is better seen in the He II spectrum (right panel) and it 
could be explained by the bonding of oxygen with silicon, leaving some charge 
available to be transferred to the titanium substrate. 
This charge is partially connected to the occupation of the empty 3d levels 
of the Ti$^{4+}$ to give Ti$^{3+}$, as reflected in Ti 2p XPS and UPS spectra, 
and the remaining charge is distributed in the available conduction band
(e.g., as seen in our calculated DOS, where empty states mainly associated to
Ti-5f are filled), inducing a shift in the valence band seen in our experiments. 
At 3 ML of Si coverage, a broad non-structured feature dominates the spectra 
between 4 and 9 eV below the Fermi level, and it is assigned to the non-bonding 
O 2p orbital of the \ce{SiO2} valence band.\cite{46} 
The line shape of the spectrum resembles the ones reported in the literature 
for thin films of amorphous silicon oxide.\cite{schoeder02} 
This suggests a disordered layer, similar to the case of Si on the 
$1 \times 2$ reconstructed surface.\cite{abad06} 
Notice that there is a slight emission increase in the band gap region between 
0 and 3 eV below de Fermi energy (inset  in Fig.~\ref{fgr:6}). 
This increase is not only due to Ti 3d states being populated, as it is shown by XPS in Fig.~\ref{fgr:5}. 
These states are located approximately at 0.7 eV binding energy and they do not spread beyond 1.0 eV from the Fermi level. 
We should mention that the presence of a SiOx layer might induce
a shift in these states.
However, we do not observe peaks in
the band gap, only a constant increase in emission from the
Fermi level to the minimum of the valence band at 3 eV 
(see inset in Fig.~\ref{fgr:6}). 
Therefore, we favor the interpretation that this is due to the contribution of 
Si states appearing in the gap, an option supported by the theoretical DOS 
calculations (see below).
Also, a shift of the valence band is observed, which moves away 
from the Fermi level between 0.3--0.4 eV. 
This effect is better appreciated in the He II spectrum. 
In addition, a new electronic state appears in the region of 10 eV, 
which is attributed 
to O 2p --  Si 3s, 3p bonds.\cite{schoeder02} 
Both for low and high coverages a decrease of the work function 
of $0.3 \pm 0.1$ eV is seen. 
For the low coverage of 0.3 ML on the (1x2) reconstructed surface an 
identical variation of the work function has been measured.\cite{abad06} 
Nevertheless, at higher coverage the decrease of the work function is larger on the (1x2) reconstructed surface. 
This difference might be related to the ordered surface defects
found in the (1x2) reconstruction, but not in the 
stoichiometric (1x1) surface.\cite{abad04,50} 
These observations are confirmed by our DFT calculations. 
In Fig.~\ref{fgr:7} we show the calculated total DOS of the clean 
\ce{TiO2} (110)-(1x1) surface compared with 
the corresponding 1 and 3 Si-atoms reconstructions ((a) and (b) respectively). 
While in the clean surface there is a clear gap of around 2.5 eV,
new peaks appear in the band gap when Si adatoms are included in the 
calculations turning the original insulator system into a metallic one
(the number of peaks increasing with the number of Si-atoms adsorbed).
This is in good agreement with the experimental result we show in the 
inset of Fig.~\ref{fgr:6} where new states appear in the band gap 
after the Si deposition. 
Fig.~\ref{fgr:8} shows a comparison between the total 
DOS for the three Si-atoms structure 
and the experimental UPS data for the 3 ML case. 
To facilitate the visual comparison between 
the experimental UPS and the theoretical DOS 
we have linear-subtracted the experimental background,
aligned theory and experiment to a common Fermi level,
and normalized the experimental and theoretical valence band
maxima to the same value (arbitrary units).
The agreement is good, in particular the valence band width is very similar. 
The experimental result, however, 
shows less structure than the theoretical one;
this could be associated with the finite temperature smearing out
theoretical peaks computed for T=0 K and the existence of disorder.
In this figure we also show the atomic contribution 
to the DOS of the most important atoms 
in the surface in the three Si-atoms optimum configuration: 
the Si-atom, the O-2f atom and both Ti-atoms. 
Due to the charge redistribution on the surface, 
originally empty Ti-states are partially filled , 
as it is found experimentally in XPS Ti 2p (see Fig.~\ref{fgr:5}).
Another interesting point is related  
to the lower part of the DOS, between -9 and -11 eV. 
Around this energy some additional peaks appear in the Si/\ce{TiO2}-DOS. 
We can see in Fig.~\ref{fgr:8} how these peaks are related to the O-Si bonds, 
as we have previously discussed in the discussion of UPS results 
(Fig.~\ref{fgr:6}).

\begin{figure}
\includegraphics[width=0.95\columnwidth]{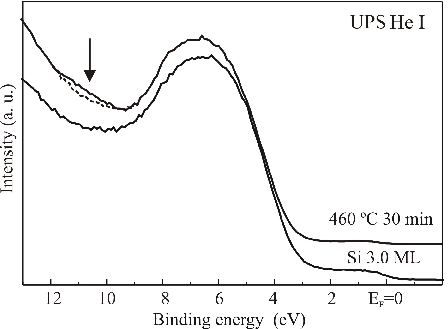}  
\caption{UPS He I spectra for a 3.0 ML Si film before and after annealing the film at $460^{\circ}$ C during 30 minutes. 
In the upper spectrum the line of points represents the background of the pristine film to stress 
the presence of new states in the region of 10-12 eV}
  \label{fgr:9}
\end{figure}

Fig.~\ref{fgr:9} shows the structure of the UPS valence band spectra of 
a 3 ML Si film before and 
after annealing at $460^{\circ}$C during 30 minutes. 
The spectrum shows essentially only one significant change: 
the appearance of a new structure in the 10-12 eV region. 
This feature also appears in Si films of 1 and 4 ML on  the 
$1 \times 2$ surface reconstruction after a heat treatment.\cite{abad06} 
We assign it to a strong Si-O bond between
O 2p and Si 3s, 3p.\cite{schoeder02,51} 
This is a clear indication of the presence of SiO$_{x}$ on the surface as 
shown by AES and XPS. 
The absence of  structure in the O 2p orbital (4-8 eV region) clearly 
confirms the relation between the 
shape of O 2p valence band orbitals and the surface order,\cite{schoeder02} 
since after annealing the (1x1) LEED pattern is not recovered (not shown). 

Experiments have been performed at room temperature, while all the calculations are at $T=0$ K, explaining why the experimental peaks show larger widths. Defects and disorder in the experimental system should further contribute to wash out peaks. However, there is an overall good agreement between the experimental UPS spectra and the calculated DOS that reinforces our interpretation of the data in terms of the theoretical models. 
This agreement is based in: (i) the similar width and shapes of the valence band 
shape and width in experiment and theory (Fig.~\ref{fgr:8}), 
(ii) the nice correlation between the predicted theoretical contribution 
from Si around 10-11 eV below the Fermi level (Fig.~\ref{fgr:8}) and the 
corresponding experimental features (marked with an arrow in Fig.~\ref{fgr:9}), 
and (iii) the presence of Si states in the band gap (Fig.~\ref{fgr:8}), 
agreeing with the increased emission observed in the UPS experiment for the 
Si-covered surface (inset of Fig.~\ref{fgr:6}(a)).

\section{Conclusions}
Thin Si films have been prepared on \ce{TiO2} (110)-(1x1) surface at room temperature in order to study 
the initial growth of Si overlayers on this surface. 
We conclude  that for very low coverages (0.3 ML) 
the silicon grows as \ce{SiO2} 
and suboxides SiO$_{x}$ with $1 \le x \le 2$. 
Furthermore, the substrate is reduced and the surface 
order begins to disappear. 

At higher Si coverages (3 ML) a film is formed by \ce{SiO2}, 
a mixture of suboxides SiO$_{x}$ with $0 \le x \le 2$, pure Si,
and cross-linking Ti-O-Si bonds found at the interface region. 
For this coverage the surface order already has disappeared completely. 
In addition, Si/\ce{TiO2} (110) interaction is strong and the interface is not atomically sharp. 
While the reaction is stronger for the (1x2) as compared to the (1x1) surface, the chemistry is similar in both cases:
Si wets completely the surface.\cite{abad06}
After thermal annealing at $460^{\circ}$ C during 30 minutes, the \ce{SiO2} stoichiometry is improved, 
but the surface order is not recovered. 
Our interpretation of the experimental data 
is further supported by ab-initio DFT calculations where optimum
configurations and corresponding electronic structure have been obtained. 
Comparison between theory and
experimental data allows us to confirm that the new states appearing in the gap are mostly due to the adsorbed Si-atoms 
(with some contribution from the Ti-atoms). 
New states located in the bottom of the valence band are related to the Si-O bonds formed in the system.

J.A. acknowledges financial support from the Ministerio de Educacion y Ciencia through grants 
MAT2006-12970-C02 and NAN2004-09183-C10-3. 
C.G. acknowledges the CSIC JAE-DOC contract under the program 
{\it Junta para la Ampliacion de Estudios} co-funded by FSE, and Spanish MEC 
under grants MAT-2008-01497/NAN and CSD2007-00041.


\newpage

\begin{table}
\caption{Size and spatial cutoffs, $R_{c}$ ({\AA}), defining the basis set used to compute
DOS projected on different atoms. 
}
\begin{tabular}{lllll}
element & basis & s & p & d \\\hline
O           & double & 1.75 & 2.01 & \\\hline
Si           & single & 2.54 & 2.86 & \\\hline
Ti           & single & 3.28  & 3.54 & 3.02 \\\hline
\end{tabular}
\label{tbl:1}
\end{table}

\begin{table}
\caption{
For different Si coverages and the optimum configuration, 
Si-O and Si-Si distances (in {\AA}), and
Si Mulliken populations (Q, in units of the electron charge). 
Labels n and m refer to Si atoms as in Fig.\ref{fgr:3Si}.
}
\begin{tabular}{cll}
$\theta$ &  & \\\hline
1 Si / 1x1 &
   \begin{tabular}{llll}
   Q       & n & Si-O  & Si-Si  \\\hline
   +0.51 & 1 & 1.65  & \\
   \end{tabular} \\\hline
2 Si / 1x1 &
   \begin{tabular}{llll}
    Q       & n & Si-O  & Si-Si \\\hline
    +0.44 & 1 & 1.70 &  2.69 \\
    +0.22 & 2 & 1.82 &   \\
      \end{tabular} \\\hline   
 3 Si / 1x1 &
   \begin{tabular}{lllll}
   Q        & Si-O & n & Si-Si & m  \\\hline
   +0.45  & 1.69& 1 &  2.77  &  2 \\
   +0.25  & 1.81 & 2 & 2.45 &   3 \\
   -0.10   & 3.73 & 3 & 2.66  &  1 \\
   \end{tabular} \\\hline
\end{tabular}
\label{tbl:2}
\end{table}

\newpage

\end{document}